\documentclass[aps,pra,twocolumn,superscriptaddress,preprintnumbers,nofootinbib,longbibliography,floatfix]{revtex4-1}
\usepackage{graphicx}
\usepackage{dcolumn}
\usepackage{bm}
\usepackage{enumitem}
\usepackage{mathtools}
\usepackage{amsfonts}
\usepackage{adjustbox}
\usepackage{float}
\restylefloat{table}
\usepackage{subcaption}
\usepackage{multirow}
\usepackage[colorlinks=true
,urlcolor=blue
,anchorcolor=blue
,citecolor=blue
,filecolor=blue
,linkcolor=blue
,menucolor=blue
,pagecolor=blue
]{hyperref}
\usepackage{braket}
\usepackage{amsmath,amsfonts}
\usepackage[table]{xcolor}

\def\begsub#1#2\endsub{\begin{subequations}\label{eq:#1}\begin{align}#2\end{align}\end{subequations}}
\def\ba#1\ea{\begin{align}#1\end{align}}
\def\bas#1\eas{\begin{align*}#1\end{align*}}

\usepackage{acronym}

\begin{document}

%
%

\title{Identifying overparameterization in Quantum Circuit Born Machines}
\thanks{This manuscript has been authored by UT-Battelle, LLC, under Contract No. DE-AC0500OR22725 with the U.S. Department of Energy. The United States Government retains and the publisher, by accepting the article for publication, acknowledges that the United States Government retains a non-exclusive, paid-up, irrevocable, world-wide license to publish or reproduce the published form of this manuscript, or allow others to do so, for the United States Government purposes. The Department of Energy will provide public access to these results of federally sponsored research in accordance with the DOE Public Access Plan.}%

\author{Andrea Delgado}
\email{delgadoa@ornl.gov}
\affiliation{Physics Division, Oak Ridge National Laboratory, Oak Ridge, TN 37831}

\author{Francisco Rios}
\email{riosfr@ornl.gov}
\affiliation{Computational Sciences and Engineering Division, Oak Ridge National Laboratory, Oak Ridge, TN 37831}

\author{Kathleen E. Hamilton}
\email{hamiltonke@ornl.gov}
\affiliation{Computational Sciences and Engineering Division, Oak Ridge National Laboratory, Oak Ridge, TN 37831}

\begin{abstract}
In machine learning, overparameterization is associated with qualitative changes in the empirical risk landscape, which can lead to more efficient training dynamics. For many parameterized models used in statistical learning, there exists a critical number of parameters, or model size, above which the model is constructed and trained in the overparameterized regime.  There are many characteristics of overparameterized loss landscapes. The most significant is the convergence of standard gradient descent to global or local minima of low loss.  In this work, we study the onset of overparameterization transitions for quantum circuit Born machines, generative models that are trained using non-adversarial gradient-based methods.  We observe that bounds based on numerical analysis are in general good lower bounds on the overparameterization transition.  However, bounds based on the quantum circuit's algebraic structure are very loose upper bounds.  Our results indicate that fully understanding the trainability of these models remains an open question.
\end{abstract}

\maketitle


%
%
\section{Introduction}
\label{sec:intro}

Variational quantum algorithms (VQAs) are at the core of near-term, gate-based quantum computing applications. By harnessing the strengths of quantum processors and the flexibility to optimize the circuit parameters through classical optimization, VQAs have found applications in supervised and unsupervised learning, with the potential for deployment on near-term devices, often limited in connectivity. 

Many types of generative models are used in machine learning, with different approaches to training and inference. In this work, we focus on generative trained to fit a known joint distribution over observed data.  A well-trained generative model can act as a surrogate to generate synthetic data similar to previous observations.  This can be a valuable tool when the underlying process which generated a given set of data is unknown, or poorly characterized, or where generating real world data has high overhead (e.g. in high energy collider experiments).  

Quantum circuits are well-suited for this task and many constructions of quantum generative models have been explored and implemented on different quantum processors and platforms: quantum Boltzmann machines were an early application for quantum annealers \cite{amin2018boltzmann}, quantum generative adversarial networks have been deployed on gate-based platforms \cite{rudolph2022iontrap_qGAN}, and the quantum circuit Born machine (QCBM) \cite{benedetti_generative_2019,liu2018qcbm} has been used to benchmark many near-term quantum devices.  QCBMs are generative models \cite{diggle1984monte,mohamed2016learning} that implement a flexible transformation using a parameterized quantum circuit trained to map an initial quantum state to a final state to generate samples from an arbitrary distribution. The construction of the circuit does not rely on an explicit form (e.g. a Hamiltonian) to construct the model. This gives us an advantage in terms of circuit design flexibility, but several recent papers have posited that increased model flexibility and expressibility comes at a consequence of trainability. In this work we focus on the trainability of QCBMs, specifically the onset of overparameterization.  

In a prior work, \cite{delgado2022unsupervised} we demonstrated that QCBMs can be used to fit arbitrary joint distributions, successfully reproducing the marginal probability distributions, and in some cases, the variable correlation. This performance is highly dependent on the underlying parameterized quantum circuit (PQC), which is used as the QCBM:  higher density connections between qubit subsets can lead to faster training and better fitting of the marginals, but fully fitting the correlations between joint variables requires large-scale entanglement or long-range correlations across all qubits. Understanding the limitations of these models, in terms of width, depth, and impact on trainability, is highly relevant for developing useful models for real-world applications.

Gradient-based optimization of QML models is made possible using analytical quantum gradients \cite{schuld2019analytic}.  It is an appealing optimization method because of known convergence properties for convex landscapes \cite{harrow2021gradient}, and the ability to find stationary points in high-dimensional spaces.  However, for non-convex landscapes, it is difficult to ensure that gradient descent will converge to the global minimum. The flatness of the objective function landscape has a direct impact on the performance of the VQA-- an extremely flat landscape may impede trainability. Furthermore, first order methods can converge to saddle points or spurious local minima (i.e. traps). Understanding the feasibility of gradient-based training is done through the characterization of the landscape, and quantifying the proliferation of traps \cite{anschuetz2022quantum}. Studies on the relationship between expressibility and trainability have indicated that highly expressive ans\"{a}tze exhibit flatter loss landscapes~\cite{Holmes2022,larocca2022diagnosing}. 

Overparameterization has been studied in classical deep learning for many classes of models and learning rules, such as spiked tensor models \cite{Vos2019landscape}, ReLU networks \cite{safran2021effects}, and random feature models \cite{baldassi2022learning}. Overparameterized models have computational capacities larger than needed to represent the distribution of the training data ~\cite{Neyshabur2018}, this often implies that the model has far more trainable parameters than the number of training points~\cite{Zhang2021}.  A key characteristic of overparameterized networks is the proliferation of local minima which are reasonable approximations of the global minimum \cite{choromanska2015loss}. Generally, overparameterized machine learning (ML) models have been associated with a fundamental change in the loss landscape structure, counter-intuitively lower training, and improved generalization properties~\cite{Buhai2020,Kawaguchi2019,Zhu2019,dar2021farewell}. Theories for what drives this transition using concepts from soft matter \cite{spigler2018jamming} or spin glass theory \cite{baldassi2022learning} have been proposed.  

Methods and analysis from ML have been employed to understand the trainability and generalization capacity of QML models. Several studies have attempted to characterize the loss landscape of quantum neural networks (QNNs) by studying the overparameterization phenomenon, as well as the capacity \cite{wright2020capacity,abbas2021power,lewenstein2021storage,Kim2021effectiveness} and expressibility of the PQC \cite{sim2019expressibility,larocca2022diagnosing}.  There have been several approaches that predict the onset of overparameterization in terms of the dimension of the Dynamical Lie Algebra (DLA)~\cite{DAlessandro2007,Zeier2011} associated with the generators of the QNN, as well as the rank of the Quantum Fisher Information (QFI)~\cite{Larocca2021,Haug2021}. In~\cite{Larocca2021}, an overparameterized model contains more trainable parameters than the rank saturation of the QFI. In both cases, the theoretical results are verified by performing numerical simulations of models trained in an unsupervised setting using energy minimization and/or with a structure inspired by a known Hamiltonian. A separate area of work has studied the proliferation of critical points in the loss landscape \cite{anschuetz2021critical,anschuetz2022quantum,you2021exponentially}. These efforts constitute attempts to study the model capacity and explore the computational phase transition, where the model's trainability is improved. 

This work presents an empirical study on the onset of overparameterization in QCBMs. What distinguishes our study from previous works is the use of unsupervised training based on the similarity of generated probability distributions. Previous studies on overparameterization and trainability transition have focused on QNNs trained using expectation values (e.g. variational quantum eigensolvers). In Section \ref{sec:qcbm_training}, we provide details on how the models are constructed and trained. We will then present our results in Section \ref{sec:overparameterization_transition}.  Additional details and results are provided in the Appendices.

\section{Gradient-based training of QCBMs}
\label{sec:qcbm_training}
Many characteristics and theories describe what occurs in the overparameterized landscape. The key concept that we are investigating is the reliable success of first-order optimization methods (gradient descent) for training overparameterized QCBM models. Our study explores the trainability of overparameterized generative models using a hybrid quantum-classical workflow that implements gradient descent to train a QCBM using the Jensen-Shannon divergence (JSD),
\begin{equation}
\begin{split}
    \mathrm{JSD}(P|Q_{\theta}) &= \frac{1}{2}\left[\mathrm{KLD}\left(P \Big\vert \frac{P+Q_{\theta}}{2}\right) + \mathrm{KLD}\left(Q_{\theta} \Big\vert \frac{P+Q_{\theta}}{2}\right)\right],\\
    \mathrm{KLD}(P|Q_{\theta}) &= \sum_{i}p(x_i)\ln{\left(\frac{p(x_i)}{q_{\theta}(x_i)}\right)}.
\end{split}
\end{equation}
This is a symmetric version of the Kullback-Leibler divergence (KLD) which has been commonly used in training generative models \cite{goodfellow2014ganORIGINAL,theis2015note} and QCBMs \cite{delgado2022unsupervised}. One advantage of the JSD loss is that it is bounded $0 \leq \mathrm{JSD}(P|Q_{\theta}) \leq \ln{(2)}$. Terms that contribute large loss values are when there is a large discrepancy between $p(x_i)$ and $q_{\theta}(x_i)$. The target distribution $P$, and generated distribution $Q_{\theta}$ are defined over a discrete set of binary strings.  $Q_{\theta}$ is generated by projecting a QCBM state prepared by a PQC onto the computational basis.
 
Multiple target distributions were used to train QCBM of $n=2, 3, 4, 6, 8$ qubits. The target distributions are  defined over $2^n$ bitstrings with varying degree of sparsity, entanglement, and correlation. Given the $2^n$-length vector of the target distribution $P(x_i)$, its sparsity ($\mathbf{s}(P)$)is the relative size of the bitstring support compared to the total length,
\begin{equation}
\begin{split}
    \mathbf{P}^{>0}&=\lbrace x_i \in [P]|P(x_i)>0\rbrace\\
    \mathbf{s}(P) &= \frac{|\mathbf{P}^{>0}|}{2^n}.
\end{split}
\end{equation}  
We use a uniform distribution (Uniform), a correlated two-dimensional joint distribution (HEP), a random sparse distribution (Sparse) and projections of the Bell state, GHZ state, and W state onto the computational basis (see Table \ref{table:targets}). The Uniform and HEP targets are dense, each bitstring has non-zero probability amplitude.  The Sparse, Bell, GHZ and W state targets are sparse, each has a finite subset of bitstrings with non-zero probability amplitudes. The Sparse, and Bell state were only used to train $2$-qubit QCBMs, the W state target was only used to train $3$-qubit QCBMs, the GHZ state target was used to train QCBMs with $n \geq 3$, and the HEP target was only used to train QCBMs with $n\geq4$. The Uniform, Sparse, Bell, GHZ and W distributions were constructed using exact values for the probability amplitudes. The HEP distribution was constructed from approximately 4 million samples drawn from the kinematic distributions of the leading jet in a simulation of the production of pairs of jets in $pp$ interactions at the Large Hadron Collider (see \cite{delgado2022unsupervised} for further detail).

\begin{table}[htbp]
\centering
\renewcommand{\arraystretch}{1.5}
 \begin{tabular}{|c|c|c|p{3.0cm}|} 
 \hline
 Target & $\mathbf{s}(P)$ & Construction & Description\\
 \hline\hline
  Uniform & 1 & Exact & $p(x_i) = 1/2^n$ for each bitstrings\\
 \hline
 Sparse & $\frac{1}{2}$ & Exact & $p(x_i) = 1/2^{n/2}$ for $2^{n/2}$ randomly selected bitstrings\\
 \hline
 GHZ, Bell & $\frac{1}{2^{n-1}}$ & Exact & $p(x_i)=1/2$ for $x_i\in [\mathrm{00..0},  \mathrm{11..1}]$\\
 \hline
W & $\frac{3}{8}$ &Exact & $p(x_i)=1/3$ for $x_i \in [\mathrm{001},\mathrm{010},\mathrm{100}]$\\ 
 \hline
 HEP &  1 & Sampled & $\approx 4$M samples from kinematic distributions in di-jet system, see \cite{delgado2022unsupervised}  \\
  \hline
\end{tabular}
\caption{Target distributions}
\label{table:targets} 
\end{table}

\begin{figure}
    \centering
    \includegraphics[width=.98\linewidth]{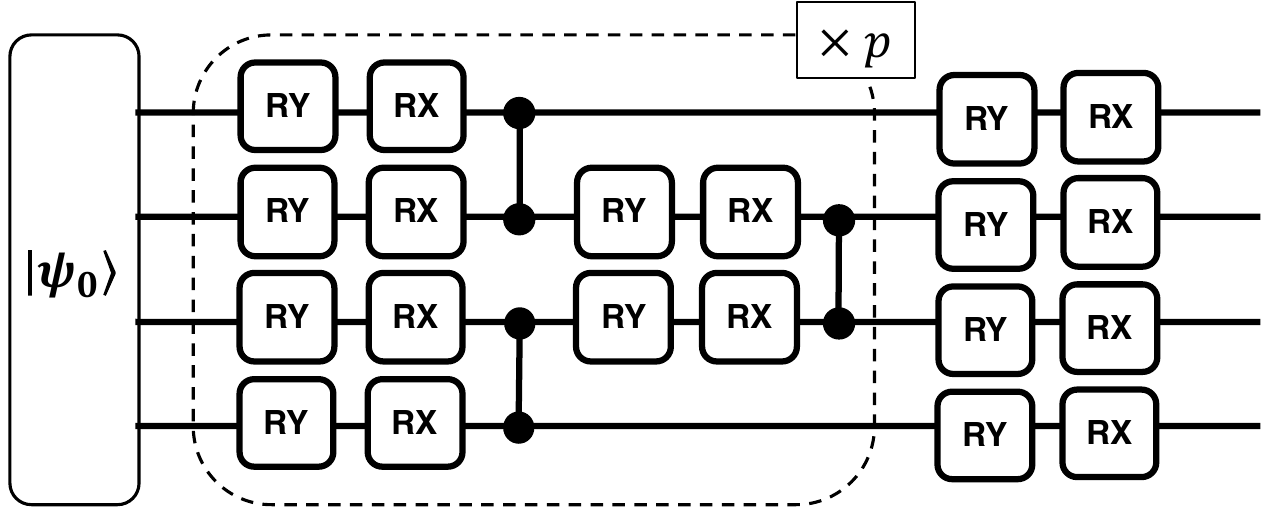}
    \caption{Example of the Hardware-Efficient Ansatz used in this study, shown with 4-qubits, open boundary conditions and acting on an initial state $|\psi_0\rangle$. The parameterized block inside the dashed boundary is repeated $p$ times for a circuit with $p$ layers.  A final layer of parameterized rotations is appended afterwards.  Measurements are made on all qubits to generate a distribution over $2^n$ bitstrings.}
    \label{fig:qcbm_circuit}
\end{figure}

The QCBM is constructed using the Hardware Efficient Ansatz (HEA)~\cite{Kandala2017}, as shown in Fig. \ref{fig:qcbm_circuit}. The HEA is built with $n$ number of qubits and $p$ layers of alternating rotational and entangling gates. The parameter size associated with a circuit of depth $p$ is $|\theta_p| = 4(n-1)p + 2n$. The single qubit gates (R$_X$, R$_Y$) are independently parameterized -- these values are updated using gradient descent. 

The gradient descent steps are implemented with adaptive moment estimation, using the Adam optimizer \cite{kingma2014adam} with hyper-parameters $\alpha=0.01, \beta_1 = 0.9, \beta_2 = 0.999$. For each QCBM used in this study, we executed $100$ independent training runs. Each training run began with a random initialization of the parameters drawn uniformly from the range $[0, 2\pi)$ and ran for $200$ steps of Adam. During training we evaluate the QCBMs in the analytical limit of $n_s \to \infty$ shots.  In Appendix \ref{appendix:finite_shot_size}, we present results with finite shot sizes.

\section{Identification of overparameterization}
\label{sec:overparameterization_transition}
\subsection{Convergence of training}
\begin{figure}
    \centering
    \includegraphics[width=.98\linewidth]{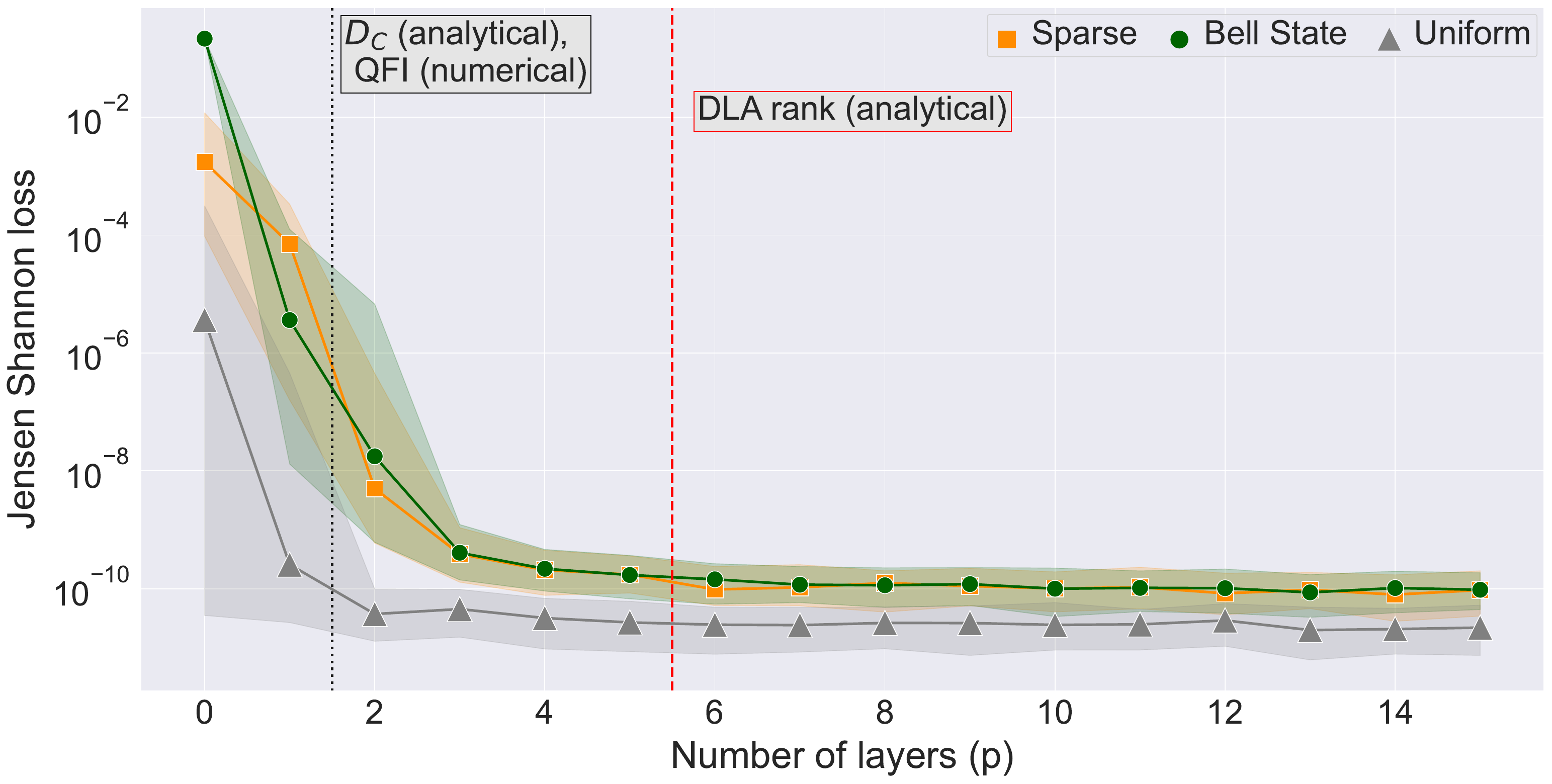}
    \caption{Median (markers) and IQR (shaded) of final losses obtained over 100 QCBM training runs ($N=2$)  as a function of the number of layers. The (dashed, red) and (dotted, black) vertical lines indicate overparameterization bounds from the literature: the $D_{C}$ \cite{Haug2021}, and QFI matrix rank saturation point \cite{Larocca2021}, and the rank of the Dynamical Lie Algebra for the HEA \cite{Larocca2021}.}
    \label{fig:pdependence2qubit}
\end{figure}

\begin{figure}
    \centering
    \includegraphics[width=.98\linewidth]{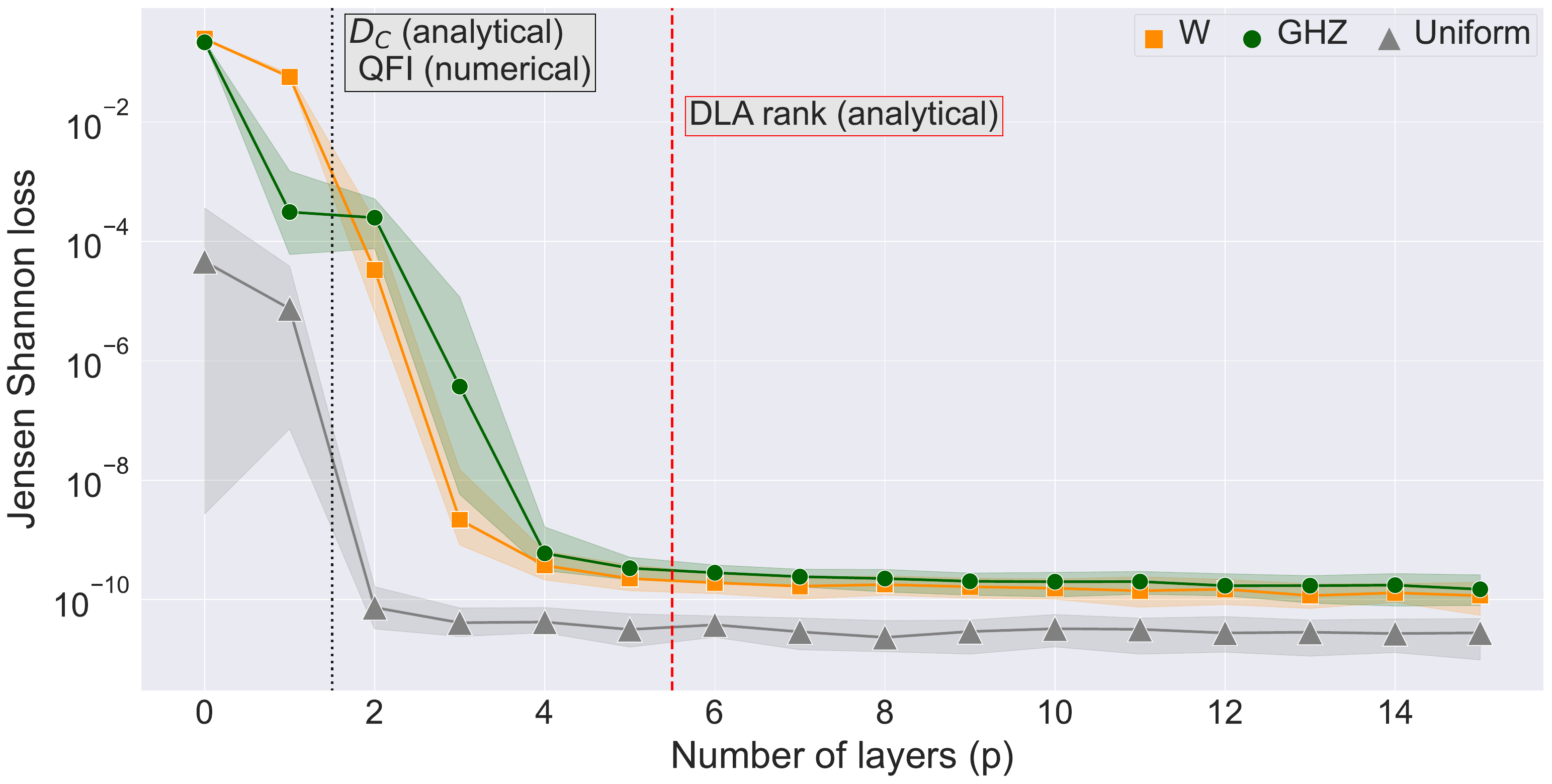}
    \caption{Median (markers) and IQR (shaded) of final losses obtained over 100 QCBM training runs($N=3$)  as a function of the number of layers. Dashed and dotted vertical lines correspond to overparameterization bounds found in the literature (see Fig. \ref{fig:pdependence2qubit}).}
    \label{fig:pdependence3qubit}
\end{figure}

\begin{figure}
    \centering
    \includegraphics[width=.98\linewidth]{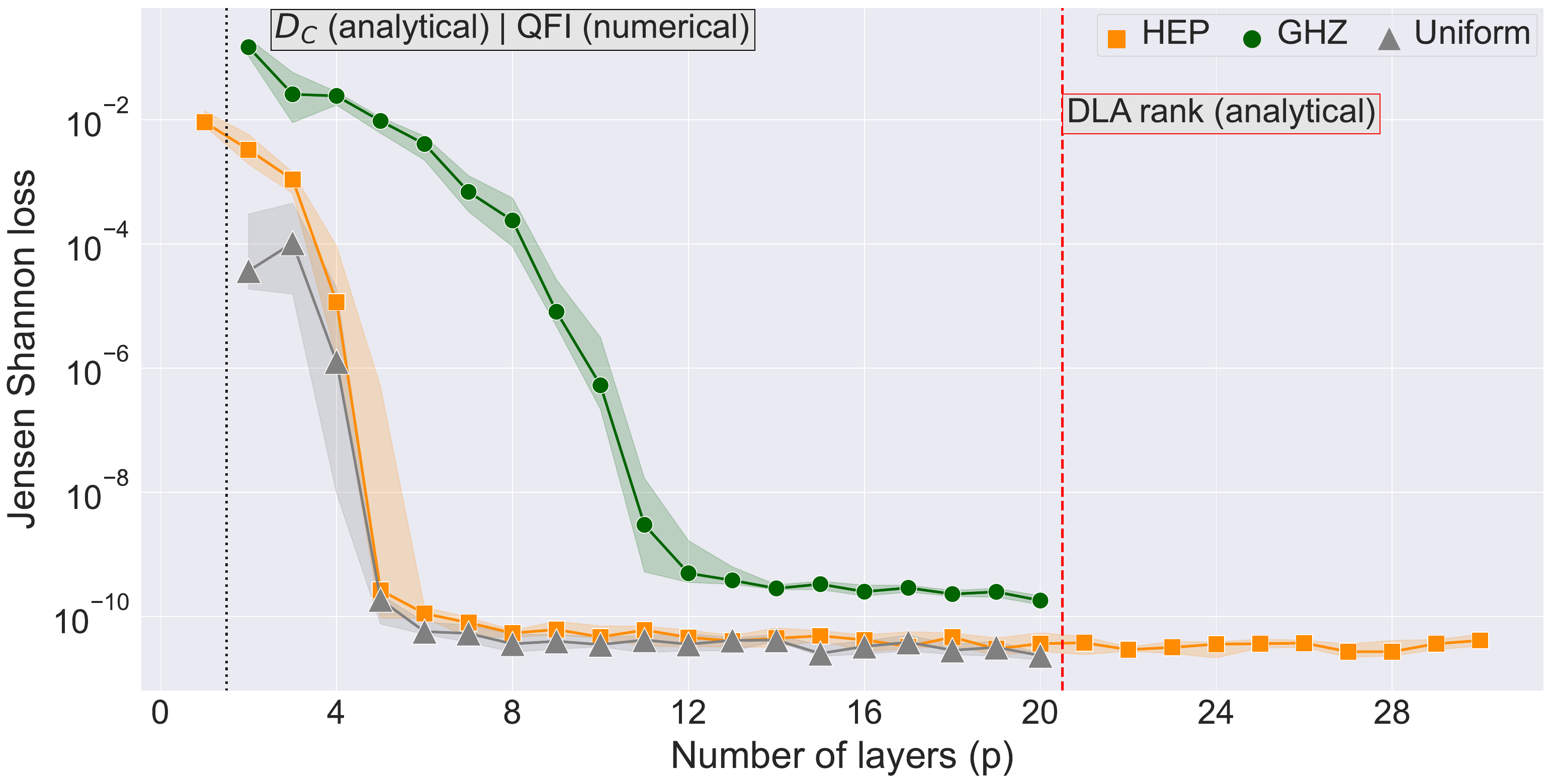}
    \caption{Median (markers) and IQR (shaded) of final losses obtained over 100 QCBM training runs($N=4$)  as a function of the number of layers. Dashed and dotted vertical lines correspond to overparameterization bounds found in the literature (see Fig. \ref{fig:pdependence2qubit}).}
    \label{fig:pdependence4qubit}
\end{figure}

\begin{figure}
    \centering
    \includegraphics[width=.98\linewidth]{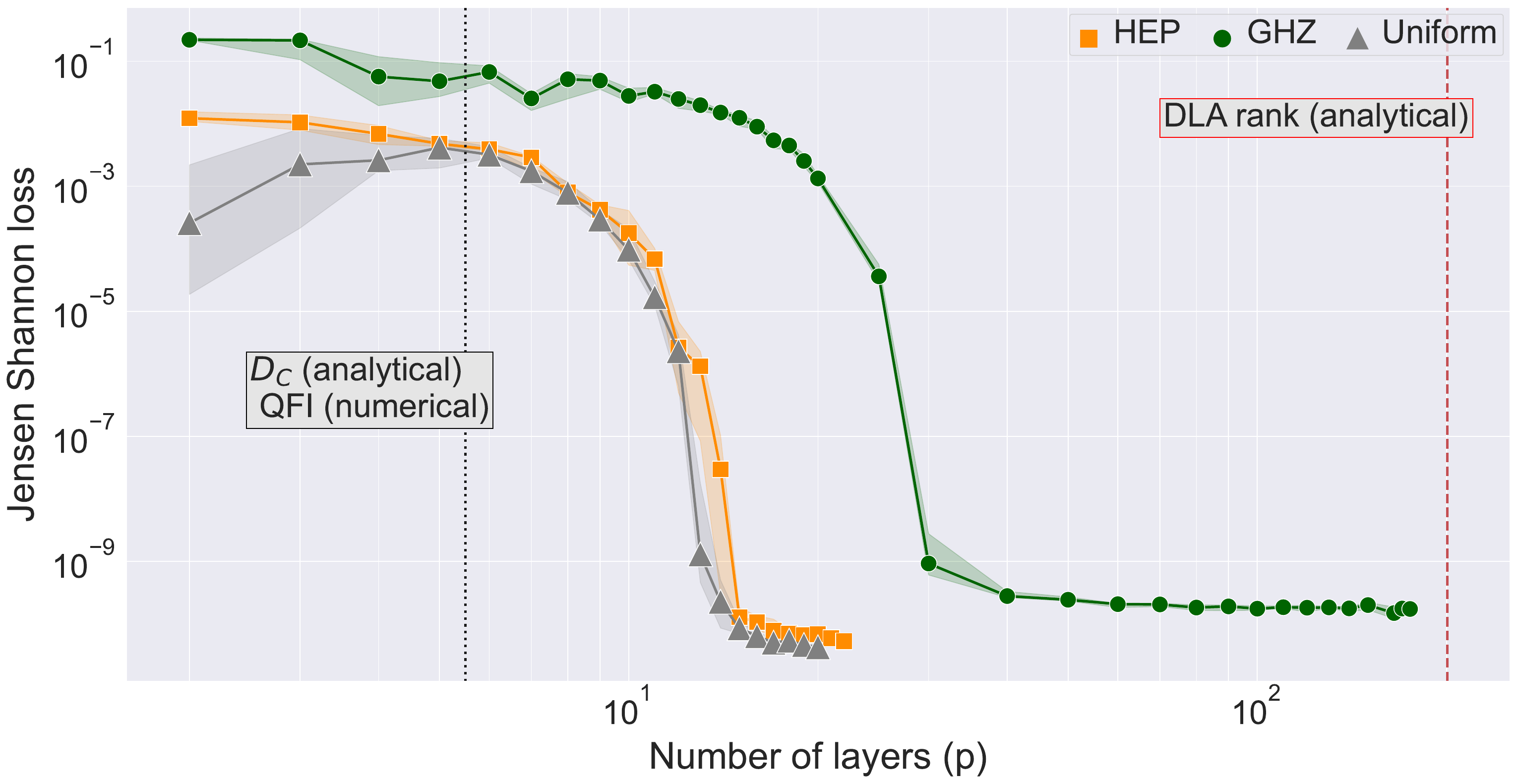}
    \caption{Median (markers) and IQR (shaded) of final losses obtained over 100 QCBM training runs ($N=6$)  as a function of the number of layers. Dashed and dotted vertical lines correspond to overparameterization bounds found in the literature (see Fig. \ref{fig:pdependence2qubit}).}
    \label{fig:pdependence6qubit}
\end{figure}

\begin{figure}
    \centering
    \includegraphics[width=.98\linewidth]{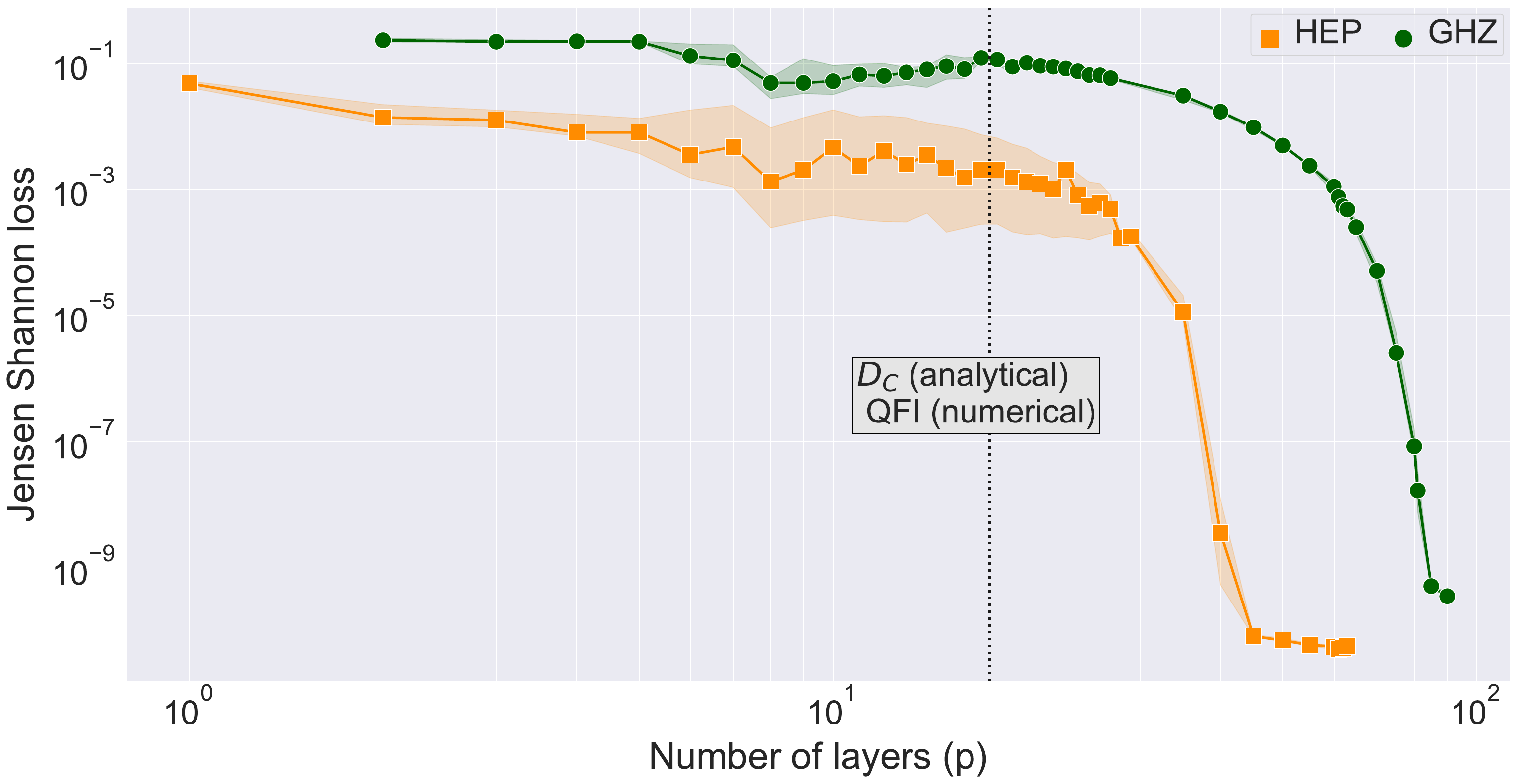}
    \caption{Median (markers) and IQR (shaded) of final losses obtained over 100 QCBM training runs ($N=8$)  as a function of the number of layers. Dashed vertical lines correspond to the $D_{C}$, and QFI matrix rank saturation point. DLA rank not shown, occurs at $p>2,340$.}
    \label{fig:pdependence8qubit}
\end{figure}

We present our results to highlight a key characteristic of overparameterized landscapes:  the reliable convergence of first-order optimization methods to low-loss regions of the landscape.  This has been observed in many different ML and non-convex optimization workflows \cite{baldassi2022learning,choromanska2015loss,ros2019landscape,safran2021effects,spigler2018jamming}. To investigate reliability in QCBM training, we train each QCBM model from $100$ randomly chosen parameter sets.  In Figs. \ref{fig:pdependence2qubit}, \ref{fig:pdependence3qubit},\ref{fig:pdependence4qubit},\ref{fig:pdependence6qubit}, and \ref{fig:pdependence8qubit} we plot the median value of the final loss value obtained from each training run. To account for the statistical spread in the data, we also shade the inter-quartile region (IQR), indicating the spread in the middle 50 \% of our obtained loss values.

In a rugged landscape dominated by traps, we expect the training to vary widely based on the initialization. The optimization may converge to low-loss solutions or it may converge to higher (poorer) minima.  This is seen in the separation between quartiles, indicated by the IQR.  We identify a critical depth $p_c$ of the QCBM where the final loss shows a sharp decrease.   

From these plots, we identify $p_c$ and report these values in Table \ref{table:phase_transitions}. For $p\ll p_c$ it is difficult for the training to escape minima of high loss (on the order of $10^{-2}$), and there is low variance in the final loss.  For $p \approx p_c$ the training becomes unstable, indicating that the quality of the found minima is highly dependent on the initialization. Once $p\gg p_c$, then each gradient descent workflow converge to losses below $10^{-8}$. The location of $p_c$ depends on the target distribution. For the Uniform distribution, it is possible to retrieve low-loss solutions with trivial $p=0$ circuits.  

\begin{table}[htbp]
\centering
\renewcommand{\arraystretch}{1.5}
 \begin{tabular}{|c|c|c|c|c|c|} 
 \hline
 \textbf{$n_{qubits}$} & \textbf{$|\psi_0\rangle$} & \textbf{target} & $p_c$(obs) & rank(QFI)\\
 \hline\hline
    2 & $|0\rangle$ &  $|\mathrm{Bell}\rangle$ & 2 & 6\\
 \hline
   2 & $|0\rangle$ &  Sparse & 2 & 6\\
 \hline
   2 & $|0\rangle$ & Uniform & 1 & 6\\
 \hline
 \hline
  3 & $|0\rangle$ &  $|W\rangle$ & 3 & 14\\
 \hline
   3 & $|0\rangle$ &  $|\mathrm{GHZ}\rangle$ & 3 & 14\\
 \hline
   3 & $|0\rangle$ & Uniform & 1 & 14\\
 \hline
 \hline
 4 & $|0\rangle$ &  $|\mathrm{GHZ}\rangle$ & 11  & 30\\
 \hline
  4 & $|0\rangle$ &  HEP & 5  & 30\\
 \hline
  4 & $|0\rangle$ &  Uniform & 4  & 30\\
 \hline
  \hline
 6 & $|0\rangle$ &  $|\mathrm{GHZ}\rangle$ & 30 & 126\\
 \hline
  6 & $|0\rangle$ &  HEP & 14  & 126\\
 \hline
  6 & $|0\rangle$ &  Uniform & 14 & 126\\
 \hline
\hline
 8 & $|0\rangle$ &  $|\mathrm{GHZ}\rangle$ & 85  & 510 \\
 \hline
 8 & $|0\rangle$ &  HEP & 40  & 510\\
  \hline
\end{tabular}
\caption{Empirical observations of critical layer depth for phase transitions in training landscape and numerically computed for QFI rank.}
\label{table:phase_transitions} 
\end{table}

\subsection{Time to solution}
\label{sec:ttsol}
Another key characteristic of overparameterized landscapes is that gradient descent efficiently converges to low-loss regions of the landscape. This has been studied in many different ML and non-convex optimization workflows. We define the time to solution (TTS[$\epsilon]$) as the minimum number of optimization steps needed to obtain a loss $\mathrm{JSD}(P|Q)\leq \epsilon$. During training, the loss is stored after each optimization step and the TTS[$\epsilon]$ metric is extracted from these stored values. For the models trained with exact analytical simulation, we take $\epsilon=10^{-8}$.

\begin{figure}
    \centering
    \includegraphics[width=.98\linewidth]{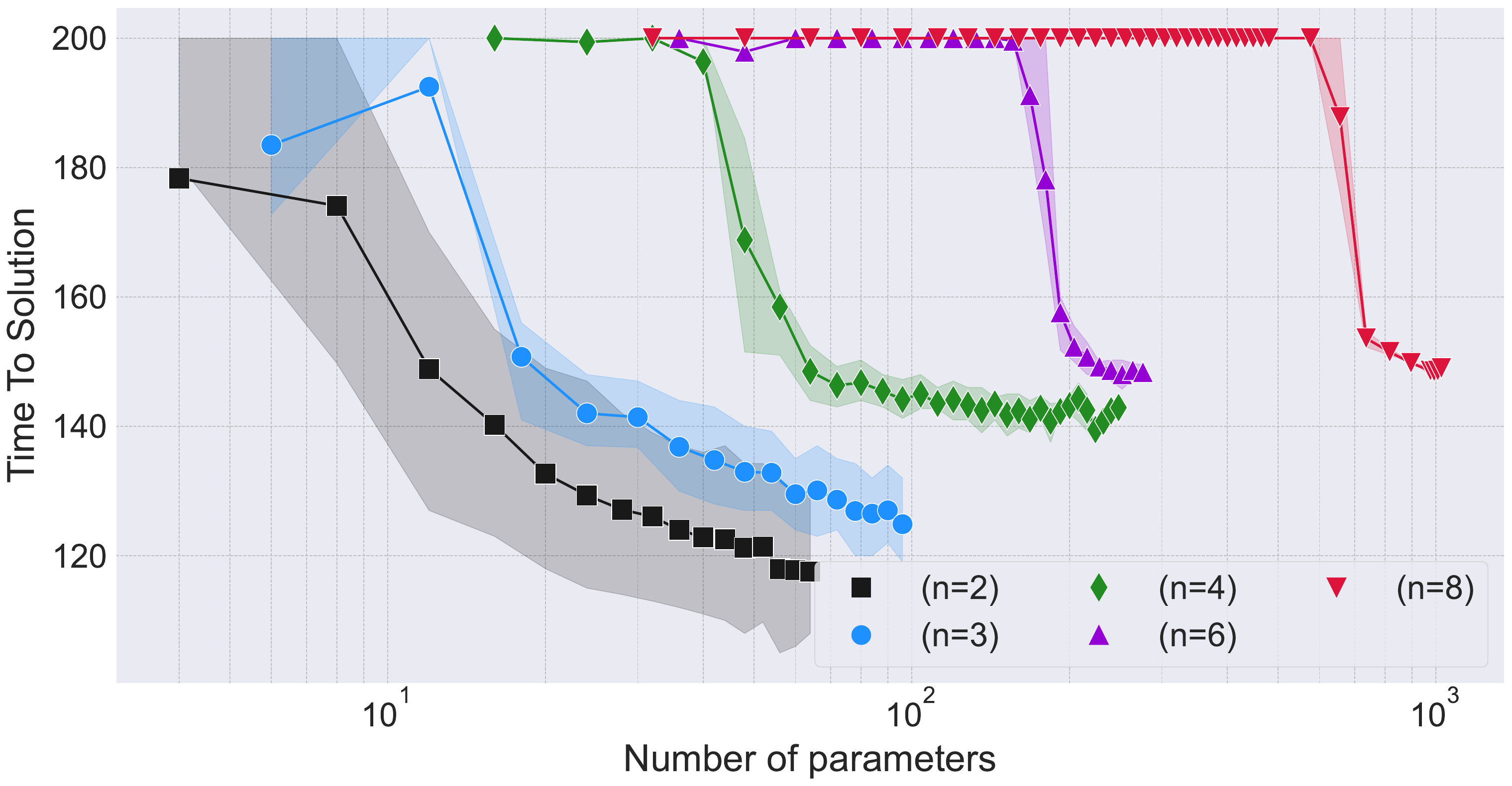}
    \caption{Median (markers) and IQR (shaded) TTS[$\epsilon=10^{-8}]$ evaluated over the Uniform, Sparse, and HEP targets used to train $2-8$ qubit QCBMs plotted as a function of the number of parameters.}
    \label{fig:ttsol_not_max_ENT}
\end{figure}

\begin{figure}
    \centering
    \includegraphics[width=.98\linewidth]{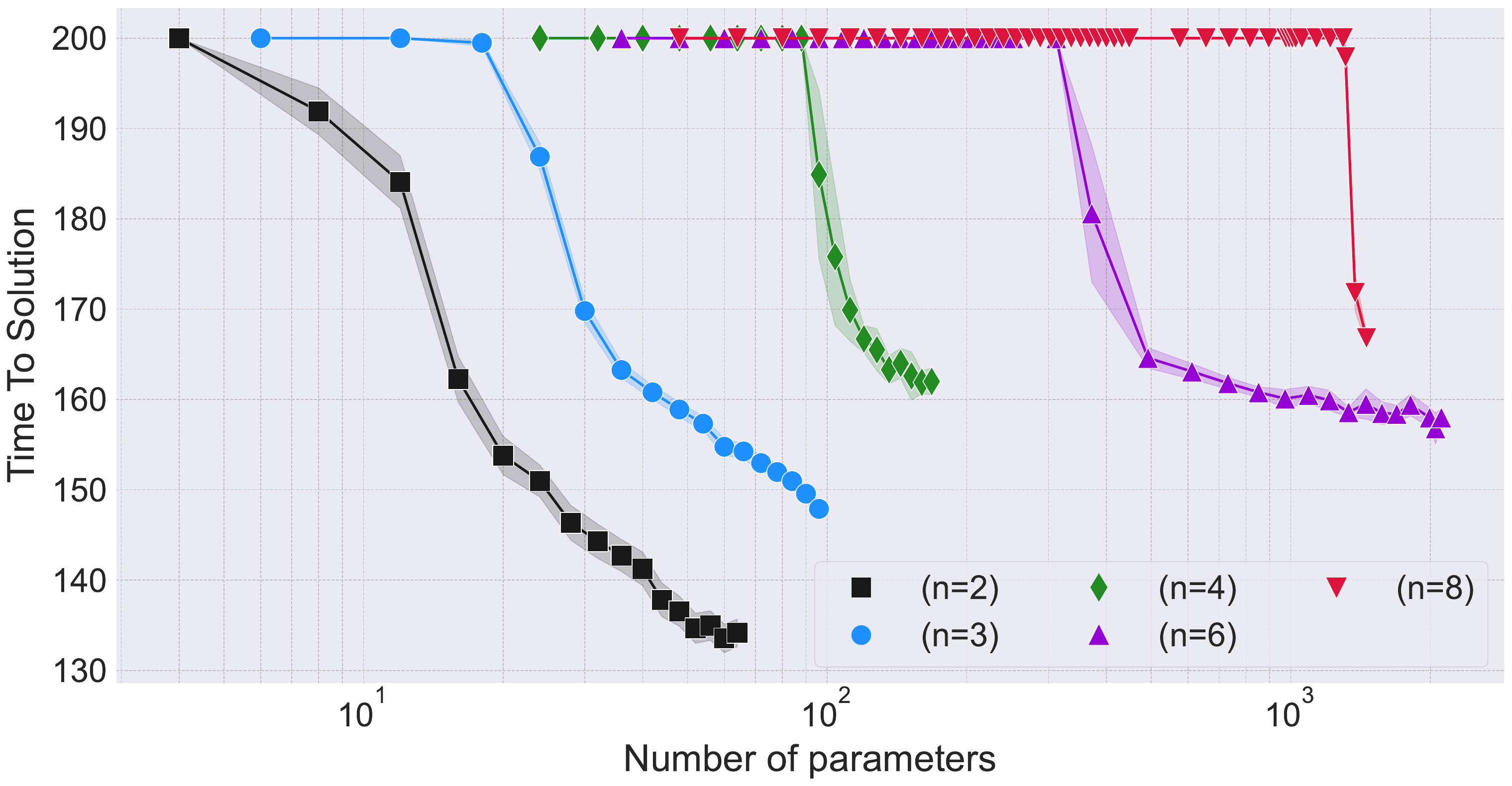}
    \caption{Median (markers) and IQR (shaded) TTS[$\epsilon=10^{-8}]$ evaluated over highly entangled targets (Bell State, GHZ, W) used to train $2-8$ qubit QCBMs plotted as a function of the number of parameters.}
    \label{fig:ttsol_maxENT}
\end{figure}

Since each QCBM was trained with a fixed number of $200$ optimization steps, if the loss function never reaches  the lower value, then $\mathrm{TTS}[\epsilon]=200$. From the values reported in Table \ref{table:phase_transitions}, we observe that overparameterization depends on the target distribution.  When reporting TTS[$\epsilon]$, we group the targets according to the amount of entanglement in the target state.  In Fig. \ref{fig:ttsol_maxENT} we plot the median TTS[$\epsilon=10^{-8}]$ for maximally entangled (Bell, GHZ, W) targets. In Fig. \ref{fig:ttsol_not_max_ENT} we plot the median TTS[$\epsilon=10^{-8}]$ for the remaining targets (Uniform, Sparse, HEP).


\subsection{Phase Transitions in Training}
\label{sec:phase_transitions}
Based on Figs. \ref{fig:pdependence2qubit}, \ref{fig:pdependence3qubit},\ref{fig:pdependence4qubit},\ref{fig:pdependence6qubit}, and \ref{fig:pdependence8qubit} there is a notable change in the QCBM training dynamics as circuit depth increases. In this section we discuss: when does a QCBM have sufficient complexity to fit the distribution, and when are low-loss solutions easy to find. When identifying these phases, we also consider their dependence on the target distribution -- we observe decreased IQR for the GHZ target, and generally for wider circuits.

First, we observe that all models trained to fit non-uniform distributions have  a minimum block depth $p>0$, needed to ensure that solutions exist. This is a direct consequence: circuits of $p=0$ can only prepare $n$-qubit product states. Additionally, it has been established that with 2- and 3-qubit states there is a minimum number of entangling gate operations that ensure a circuit can connect the entire state space \cite{perdomo2021entanglement,perdomo2021canonical,Znidari2008optimal} or prepare a maximally entangled state. Our results in Figs. \ref{fig:pdependence2qubit} and \ref{fig:pdependence3qubit} agree in the sense that, when training on the Bell state or GHZ state distribution, the circuits needed at least $p=1$ for the training to show non-trivial minimization.  At $p=1$, the training results with the Bell and GHZ state distribution can find low-loss solutions; when training on the W state distribution, an additional block is needed. With $n=4, 6, 8$ qubits, the training finds lower minima at $p=2, 3, 6$, respectively. 

Beyond these transitions, the next transition is marked by  low-loss solutions, indicating that the QCBM has sufficient complexity to fit the distribution and the training enters the overparameterization regime: low-lying solutions are reliably found for all training runs. This occurs at higher depths, which we reported in Table \ref{table:phase_transitions}.  

\section{Comparison to Quantum Bounds}
\label{sec:quantum_overparameterization}
In Table \ref{table:phase_transitions}, we reported our empirical observations of phase transitions in QCBM trainability (Figs. \ref{fig:pdependence2qubit}-\ref{fig:pdependence8qubit}) as a critical circuit size $p_c(\mathrm{obs})$, and supported this through empirical observations about time to solutions (Figs. \ref{fig:ttsol_not_max_ENT},\ref{fig:ttsol_maxENT}).  Based on these observations, we are confident that the QCBMs exhibit overparameterization. In Table \ref{table:phase_transitions}, we compare $p_c$ to several previously established bounds in the literature: the parameter dimension $D_{C}$ \cite{Haug2021}, the rank of the QFI matrix (QFI) \cite{Larocca2021}, and the dimension of the Dynamical Lie Algebra associated with the generators of the circuit ansatz ($\mathfrak{g}_{HEA})$.) \cite{Larocca2021}. The QFI rank in Table \ref{table:phase_transitions} is numerically computed, while $D_{C}$ and the DLA rank $\mathrm{DLA}=\mathrm{dim}(\mathfrak{g}_{HEA})$ are analytically evaluated. 

The expressive power of a PQC is associated with $D_{C}$.  It has a maximum possible value of $D_{C} = 2^{n+1} - 2$ and quantifies the number of independent parameters in a quantum state prepared by a PQC. It is also related to the QFI~\cite{Yamamoto2019, Stokes2020}, which considers the quantum geometric structure of the PQC. The QFI matrix rank is computed numerically, and in Appendix \ref{appendix:trainability_metrics} we provide further detail on its computation.  The rank of the QFI matrix saturates at the maximum value of $D_C$. The DLA rank $\mathrm{DLA}=\mathrm{dim}(\mathfrak{g}_{HEA})$ is an upper bound on the number of parameters in a circuit, to define a local map that takes any element from $(\mathfrak{g}_{HEA})$ to a point in parameter space.  

In Figs. \ref{fig:pdependence2qubit}--\ref{fig:pdependence8qubit}, we observe that for circuits with $N\leq 4$ qubits, $D_C$ and QFI rank are approximate lower bounds on $p_c$. For all circuit widths, $(\mathfrak{g}_{HEA})$ is a loose upper bound on $p_c$.  

\begin{table}[htbp]
\centering
\renewcommand{\arraystretch}{1.5}
 \begin{tabular}{|c|c|c|c|c|c|} 
 \hline
 \textbf{$n$} & \textbf{$D_{C}$} &\textbf{$p\: (D_C)$} & DLA  & $p$ (DLA)\\
 \hline\hline
   2 & 6 & 1 & 16 & 7\\
 \hline
  3 & 14 & 1 & 64 & 7\\
 \hline
 4 & 30 & 2  & 256 & 20 \\
 \hline
 6 & 126 & 5 & 4096 & 204\\ 
 \hline
 8 & 510 & 17 & 65536 & 2340 \\
  \hline
\end{tabular}
\caption{Analytical bounds and circuit depths ($p$) needed to reach them: $D_{C}$ (\cite{Haug2021}),  $ |\theta_p| \geq D_C$, DLA rank $\mathrm{DLA}=\mathrm{dim}(\mathfrak{g}_{HEA})$ (\cite{Larocca2021}), $|\theta_p| \geq \mathrm{dim}(\mathfrak{g}_{HEA})$.}
\label{table:dcrank_v2} 
\end{table}

\section{Conclusion}
\label{sec:conclusions}
In this work, we have presented an empirical study showing the onset of overparameterization in QCBMs.  We have provided the first demonstration of overparameterization for the QCBM and loss functions evaluated with probability distributions.  The overparameterization transition occurs at a critical depth of the QCBM circuit. We observe that for overparameterized circuits, gradient-based training becomes highly efficient and that this critical depth is dependent on the target distribution.  

For classical and quantum models, understanding and identifying what drives the overparameterization transition remains an open question and for both paradigms, overparameterization is associated with counter-intuitive behavior.  In supervised training, overparameterization leads to models for which gradient-based training converges to solutions with good generalization properties without over-fitting \cite{zhang2021understanding}. In this work we report that overparameterized QCBMs can train efficiently without impedance from barren plateaus. 

As detailed in Appendix \ref{appendix:trainability_metrics} we present further evidence that the barren plateau does not prevent trainability. However, predicting long-time training dynamics from initial behavior remains challenging and an unresolved question.  More so, a full understanding of the QCBM trainability remains an open question, for both ideal (noiseless, infinite shot limit) and noisy models. This work is a first step along these paths. In Appendix \ref{appendix:finite_shot_size} we provide results on overparameterization when using finite sampling sizes. 

Our results should serve to motivate further discussion about about the trainability of QCBMs.  First,  to obtain more robust upper bounds for the onset of overparameterization,  second to understand what are the drivers of changes in the trainability, third to determine how trainability is affected by noise, and finally to build a more complete understanding of trainability beyond the barren plateau.  

\begin{acknowledgments}
This work was partially supported by the Quantum Information Science Enabled Discovery (QuantISED) for High Energy Physics program at ORNL under FWP ERKAP61. This work was partially supported by the Laboratory Directed Research and Development Program of Oak Ridge National Laboratory, managed by UT-Battelle, LLC, for the U. S. Department of Energy. This work was partially supported as part of the ASCR Fundamental Algorithmic Research for Quantum Computing Program at Oak Ridge National Laboratory under FWP ERKJ354. This work was partially supported by the ASCR Quantum Computing Applications Teams Program at Oak Ridge National Laboratory under FWP ERKJ347. This research used birthright cloud resources of the Compute and Data Environment for Science (CADES) at the Oak Ridge National Laboratory, which is supported by the Office of Science of the U.S. Department of Energy under Contract No. DE-AC05-00OR22725.
\end{acknowledgments}

\bibliography{ref}
\appendix
\section{Robustness of Overparameterization Transition}
\label{appendix:finite_shot_size}
As stated in the Introduction, the QCBM is an implicit probabilistic model -- it is trained to fit a data distribution from a (possibly unknown) process without explicitly modeling the process itself.  It is apparent from the results presented in Section \ref{sec:phase_transitions} that the landscape undergoes a transformation which results in low loss minima being easy to find in the limit of $n_s \to \infty$ shots. 
\begin{figure}[htbp]
    \centering
    \includegraphics[width=.98\linewidth]{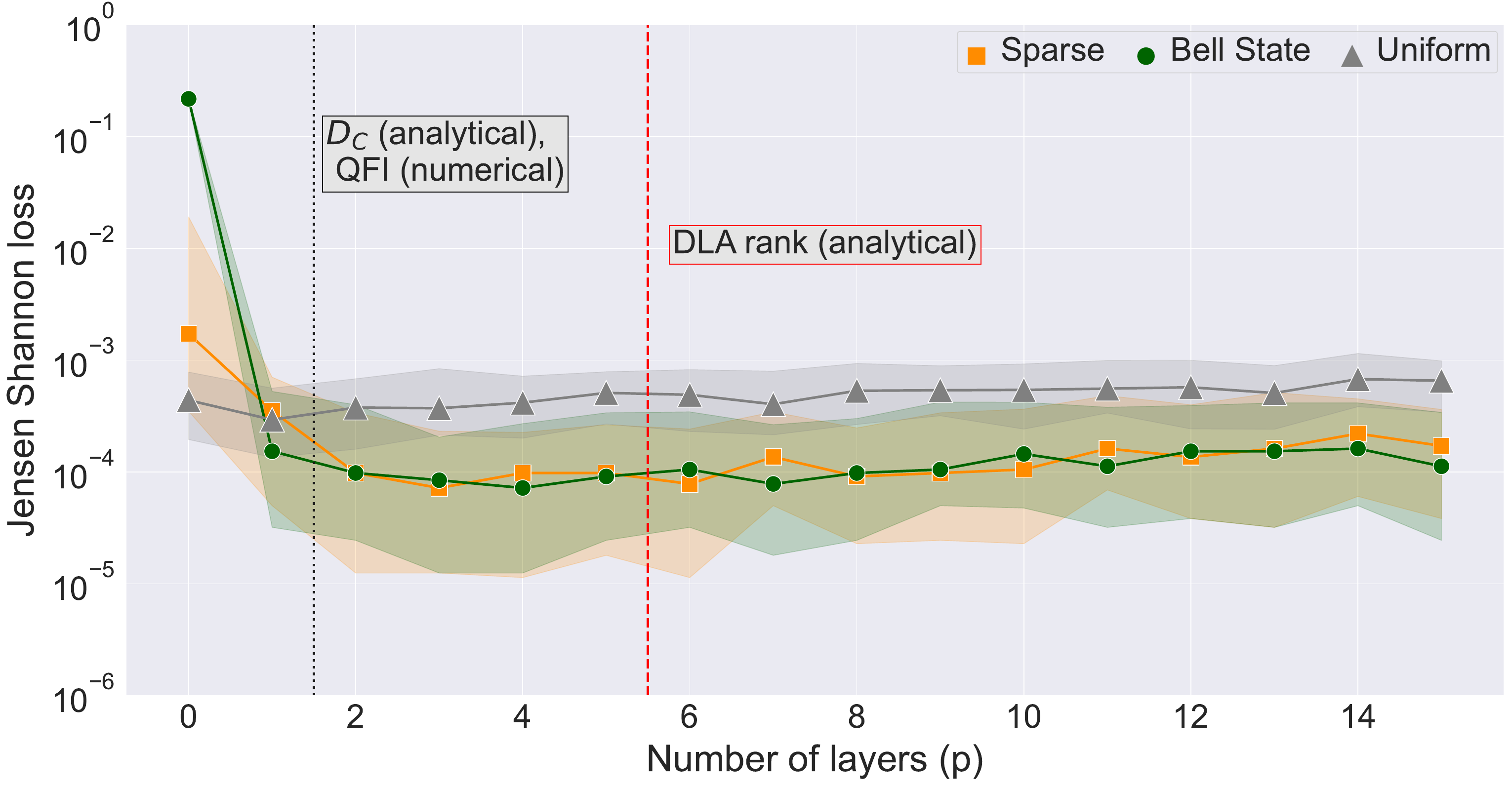}
    \caption{Median (markers) and IQR (shaded) of final losses obtained over 100 QCBM training runs ($N=2$) using finite shot size $n_s = 10^{3}$, plotted as a function of the number of layers ($p$).}
    \label{fig:pdependence2qubit_1K}
\end{figure}

The results in the main text are from QCBM trained with the exact analytical simulation of a noiseless circuit.  In this section we discuss the robustness of this transition with respect to statistical noise encountered by preparing and sampling a quantum state with finite shots.  On near-term hardware, there will be noise from many different hardware sources. As a simulation, including finite shot sizes introduces a source of statistical noise, under the assumption that the QCBM state can be prepared and measured without error, resulting in $n_s$ independent samples from the final state.    
\begin{figure}[htbp]
    \centering
    \includegraphics[width=.98\linewidth]{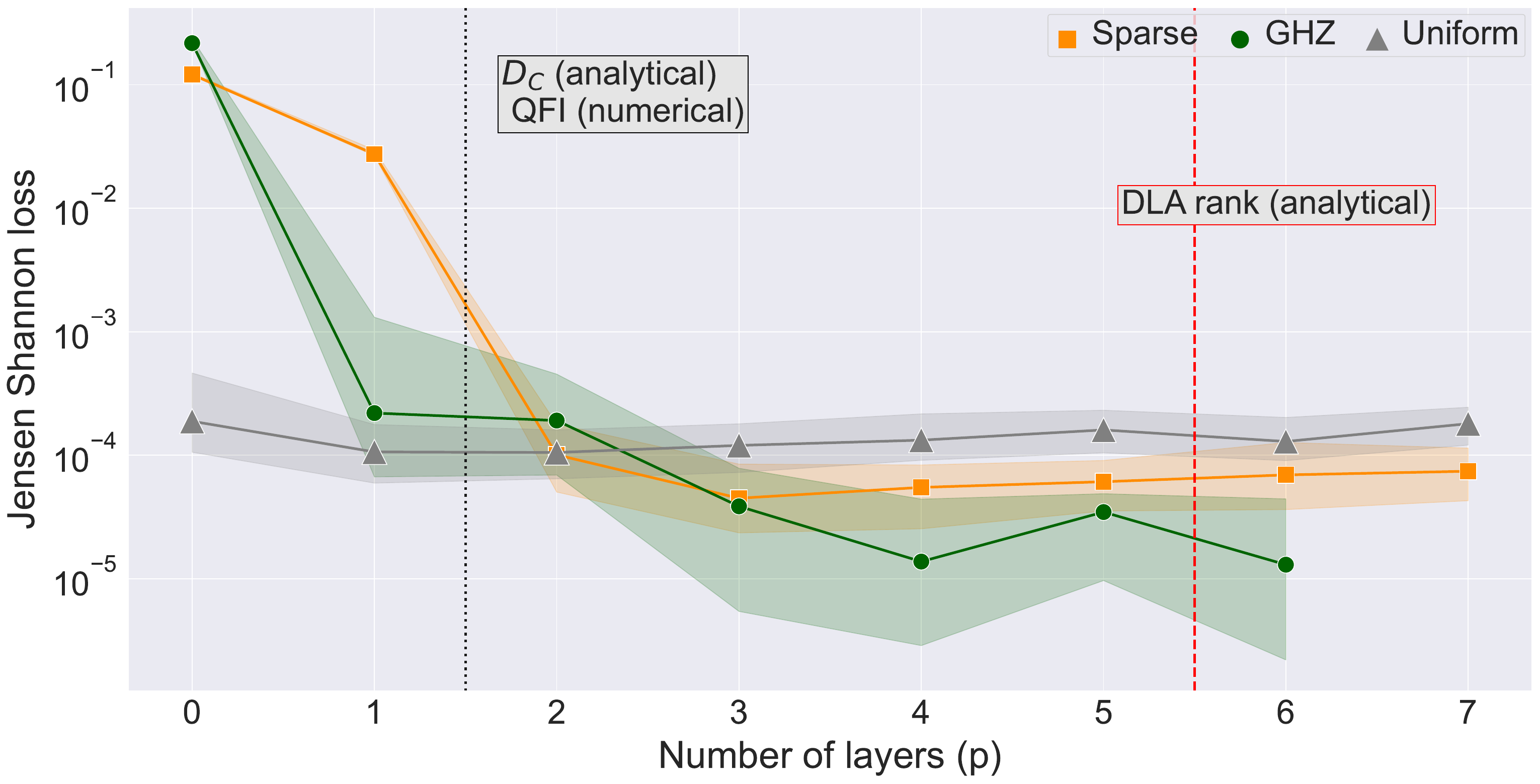}
    \caption{Median (markers) and IQR (shaded) of final losses obtained over 100 QCBM training runs ($N=3$) using finite shot size $n_s = 10^{5}$, plotted as a function of the number of layers ($p$).}
    \label{fig:pdependence3qubit_10K}
\end{figure}

In classical learning, stochastic noise is expected to act as a regularizer and also assist in escaping from local minima and avoid trapping on saddle points. Statistical noise can impede training by under-representing modalities in the target distribution, or in the sampled distribution. Thus, we investigate whether the inclusion of statistical noise affects the trainability of QCBM in particular the onset of an overparameterization transition.

With finite shot sizes, we can still observe an overparameterization transition, however the decrease in loss in the overparameterized models is not as dramatic and the loss appears to be lower bounded by $1/\sqrt{n_s}$.  Additionally, the heuristic we used to identify $p_c$ is of limited use with finite shots:  at $1000$ or $10000$ we see that the IQR does not dramatically change, even though the final loss value saturates as the model depth increases. 

\begin{figure}[htbp]
    \centering
    \includegraphics[width=.98\linewidth]{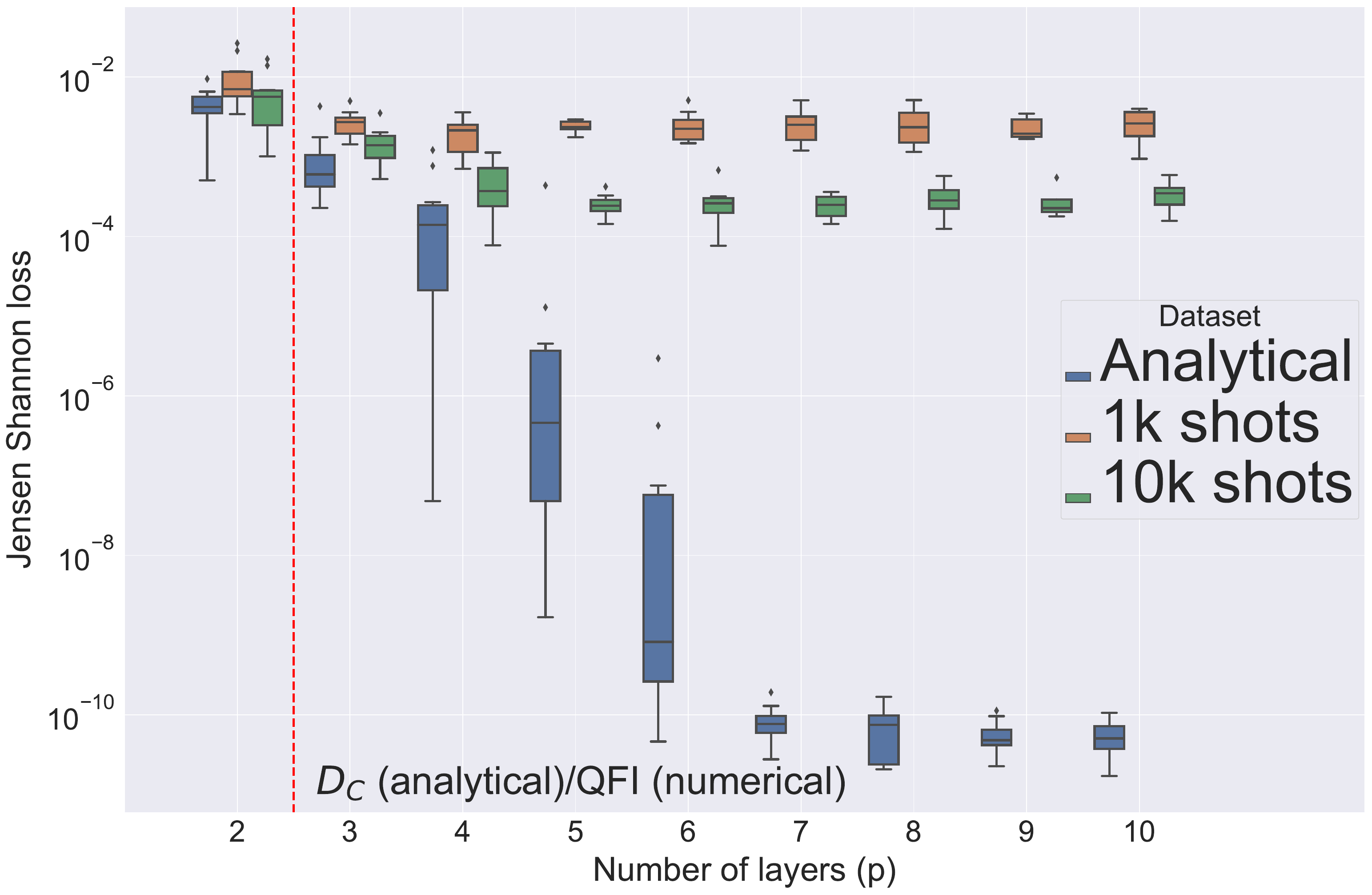}
    \caption{Quartile plots for 100 QCBM training runs using 4-qubit QCBMs built using the HEA, and trained on the HEP target.  Final loss values plotted as a function of the number of layers. Results displayed for different values of $n_s$ used during training.}
    \label{fig:pdependence4qubit_fs}
\end{figure}
\section{Predicting Trainability}
\label{appendix:trainability_metrics}
The results in the main text appear to contradict the current understanding variational quantum algorithms.  The QCBM models used in this study both exhibit overparameterization, yet the critical depths needed to reach these regions is predicted to result in models plagued by barren plateaus.  In this section we present the results for three metrics: the QFI, the variance of the gradient, and the Hessian spectrum.  These metrics have been introduced and used in the QML literature to characterize or quantify the trainability of variational quantum circuit models \cite{Larocca2021,McClean2018,huembeli2021characterizing}.  When a target is required (to evaluate the loss) we use two sample targets: the GHZ, Bell targets, and the Sparse target (see Table \ref{table:targets} in main text). The results in this section are presented are in support of our statement in Section \ref{sec:conclusions}, that a full understanding of the trainability of QCBM models remains an open question.

\subsection*{Quantum Fisher Information}
\begin{figure}[htbp]
    \centering
    \includegraphics[width=.98\linewidth]{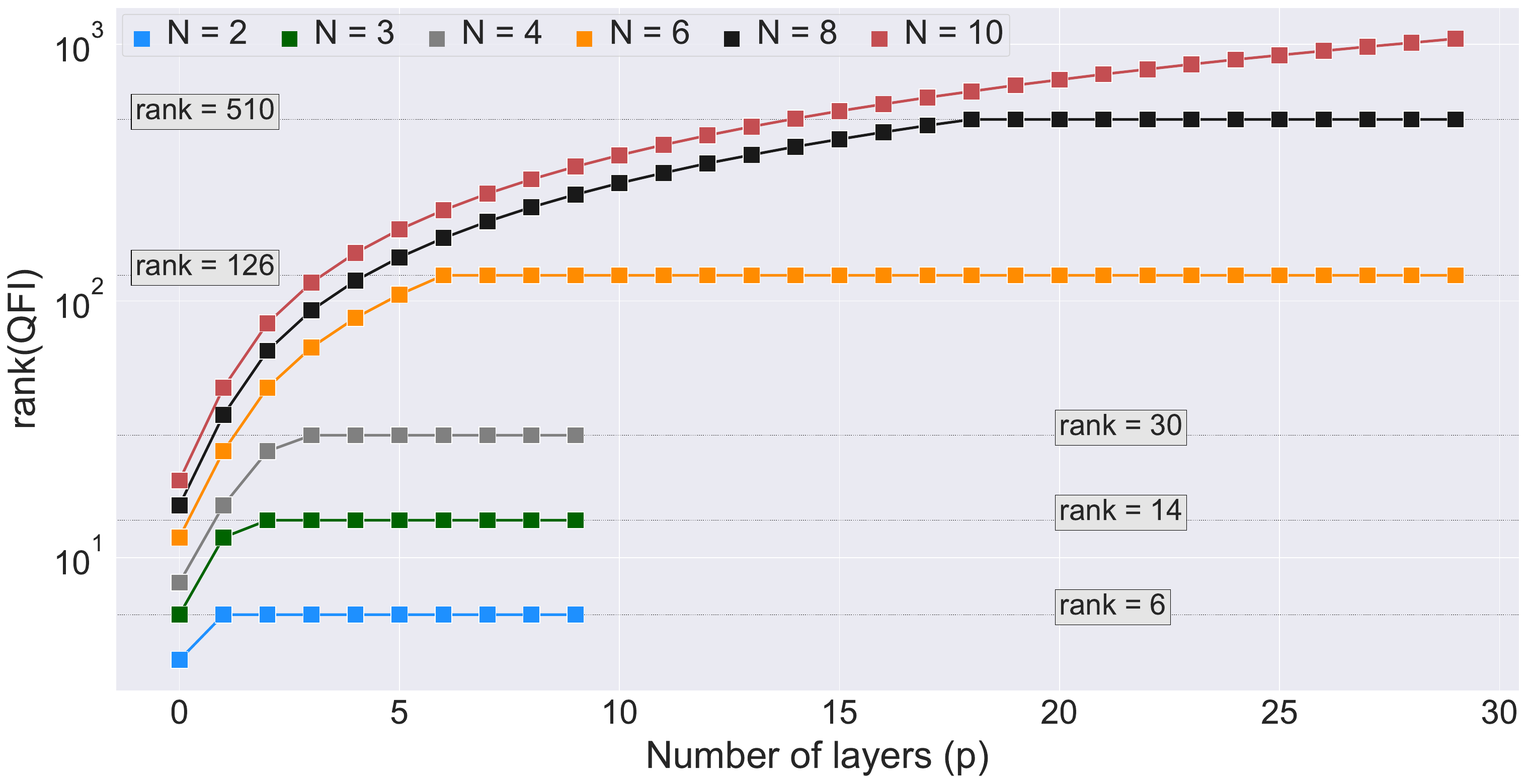}
    \caption{QFI matrix rank of QCBMs constructed with the HEA as a function of layers ($p$). Dashed black lines correspond to $D_{C}$ values reported in Table \ref{table:dcrank_v2}.}
    \label{fig:qfirank}
\end{figure}

We first compute the QFI which is a fundamental concept in quantum information. It quantifies the amount of information that a quantum state carrier about a particular parameter of interest. The Fisher information is defined in terms of the derivative of the density matrix with respect to the parameter being estimated. In the context of quantum computing, the QFI provides a measure of how sensitive the state is to changes in the parameter and is used to bound the overparameterization transition. The rank of the QFI matrix increases linearly with $p$ with a slope of $n_{qubits}$, as seen in Figure \ref{fig:qfirank}  until it reaches the maximum possible value for $D_{C} = 2^{n+1} -2$. The QFI rank is independent of the target distribution.

The saturation of the QFI matrix has been associated with the maximum number of trainable parameters in the quantum model, and it coincides with the expressive power of the circuit, or parameter dimension $D_{C}$. In Table \ref{table:phase_transitions} of the main text, we report the numerically computed rank of the QFI matrix for our QCBM models, and in Table \ref{table:dcrank_v2} we reported the analytical values of $D_{C}$. 
\subsection*{Loss Gradient Variance}
The overparameterization transition reported in the main text, occurs in circuits with circuit depths that far exceed the model sizes where barren plateaus have been predicted to exist \cite{McClean2018,Cerezo2021}.  The barren plateaus is a potential obstacle for variational quantum models where a randomly initialized circuit will sit on a plateau in the landscape with ``no interesting search directions in sight,'' (cf. \cite{McClean2018}). 

Many sources of barren plateaus have been identified from loss function design \cite{Cerezo2021}, noise channels \cite{Wang2021}, entanglement \cite{marrero2021entanglement}, to the high expressivity of the ansatz design \cite{larocca2022diagnosing}.  Additionally, barren plateaus have been used to indicate that many solutions sit in narrow ravines and can be difficult to find \cite{Holmes2022}.  
\begin{figure}[htbp]
    \centering
    \includegraphics[width=.98\linewidth]{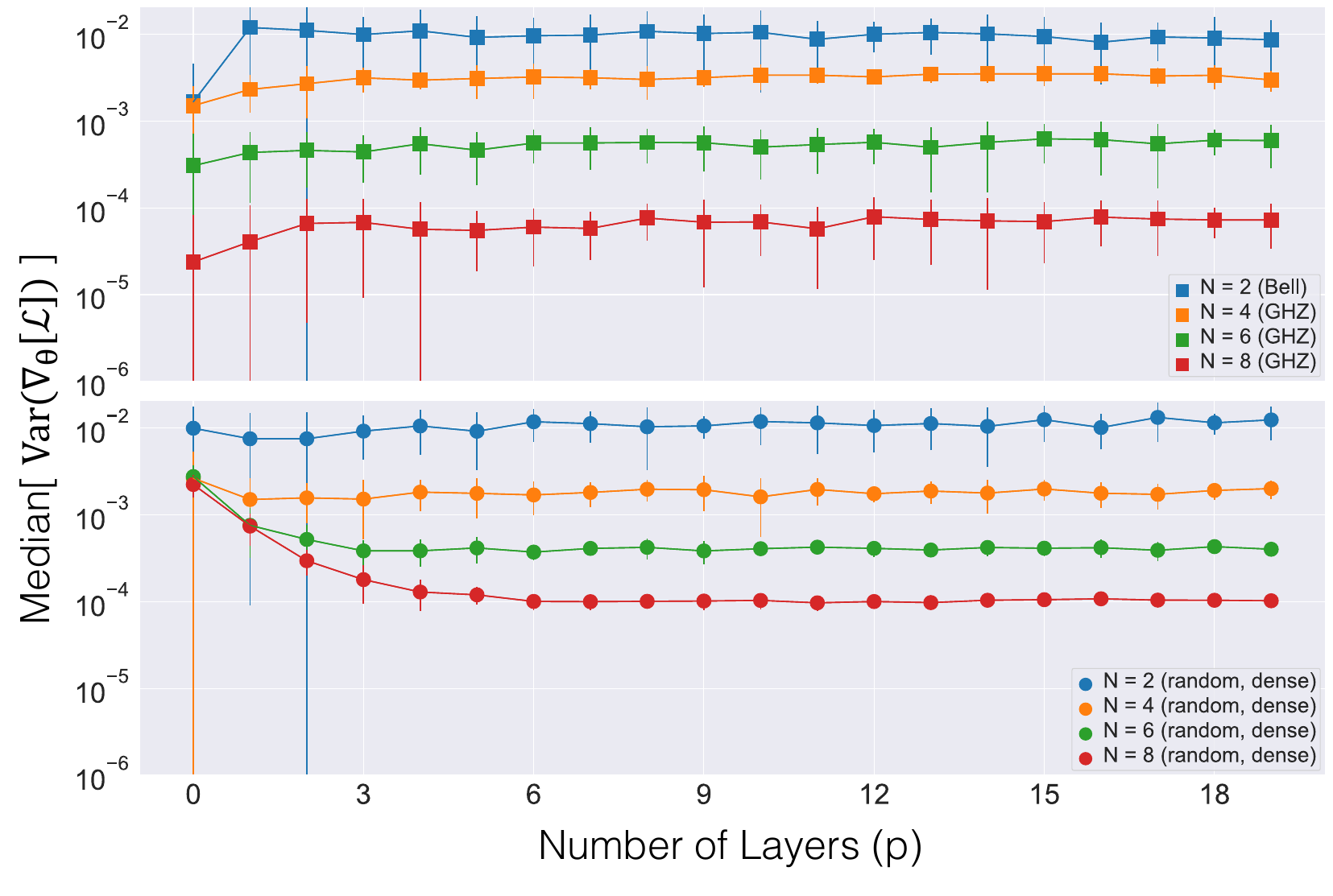}
    \caption{Median variance of the JSD gradient for QCBMs with $N = 2, 4, 6, 8$ qubits and random initializations.  Loss gradient is evaluated with respect to the Bell, GHZ targets (top) or a random target (bottom). Error bars defined by the IQR evaluated over 30 random initial parameter sets.}
    \label{fig:variance_gradient}
\end{figure}

The vanishing of all gradient directions is taken to indicate that gradient-based methods will be highly inefficient, or even fail to train. The current understanding of the QML community is that trainable models must be constructed from shallow circuits, or with restricted expressivity.  Yet in the main text we report on models built from highly expressive ansatz, constructed to depths that far exceed the $O(\mathrm{poly}(n))$ scaling.  These models have demonstrated trainability in the extreme case of $n_s \to \infty$ in the main text, but trainability also is seen for finite shot sizes.  

\begin{figure}[htbp]
    \centering
    \includegraphics[width=.98\linewidth]{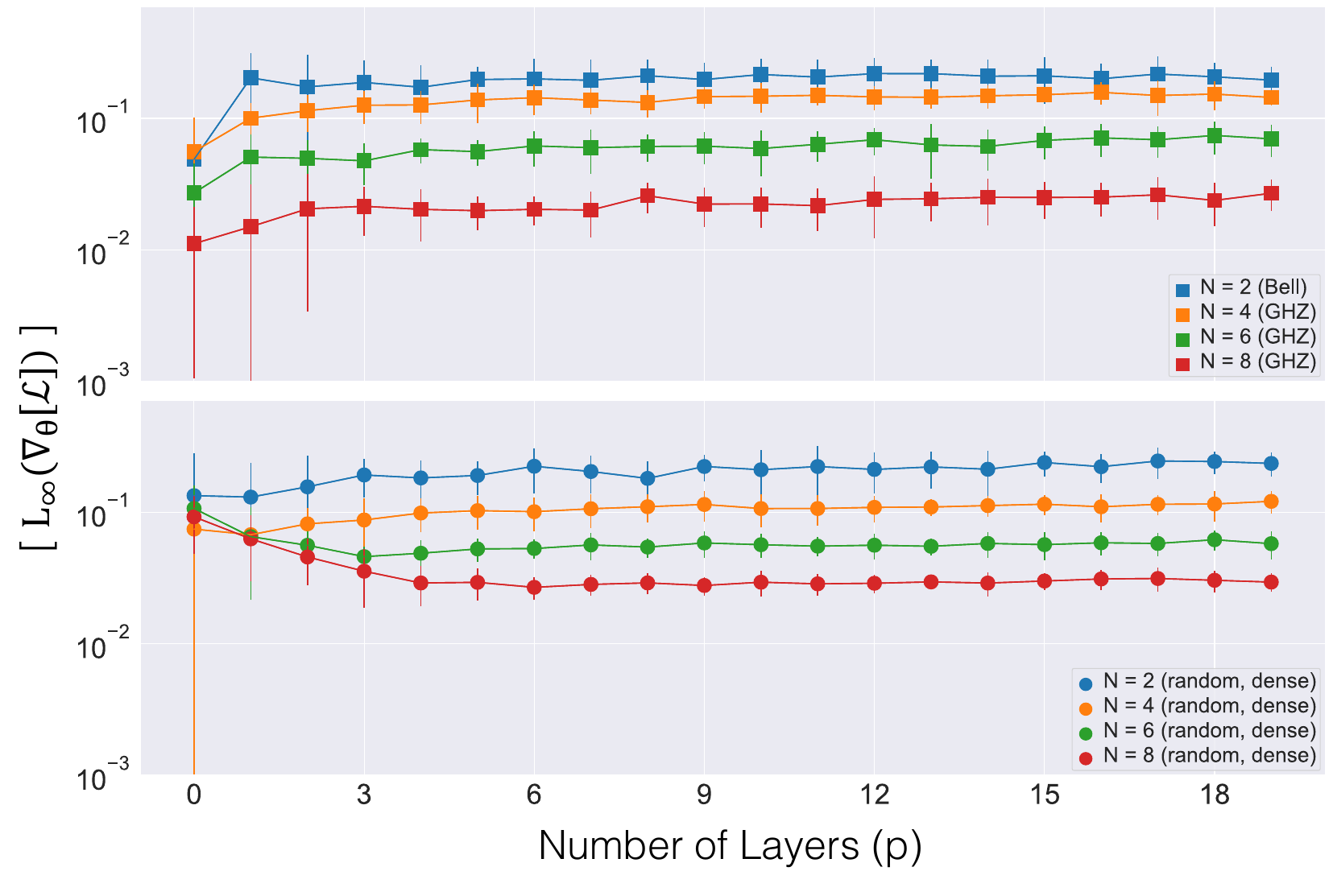}
    \caption{L$_{\infty}$-norm of the JSD gradient for QCBMs with $N = 2, 4, 6, 8$ qubits and randomly initialized.  Loss gradient is evaluated with respect to the Bell, GHZ targets (top) or a random target (bottom). Error bars defined by the IQR evaluated over 30 random initial parameter sets.}
    \label{fig:max_gradient}
\end{figure}
Identifying barren plateaus can use the variance of the loss gradient \cite{Cerezo2021}, or sample variance of a two-local Pauli term gradient \cite{McClean2018}.  We compute the variance of the loss gradient $\mathrm{Var}[\nabla_{\theta}\mathrm{JSD}(P|Q_{\theta})]$.  This variance is dependent on the choice of target, and in Fig. \ref{fig:variance_gradient} we plot the median loss variance for randomly initialized QCBMs using both targets. When evaluated against the denser target, the variance scales as predicted in \cite{McClean2018}.   However, even though the gradient variance vanishes, there remains significantly large directions (Fig. \ref{fig:max_gradient}).  

We use Figs. \ref{fig:variance_gradient}, \ref{fig:max_gradient} to emphasize that the gradient evaluated at random parameters gives an incomplete picture.  From the results in the main text, 2-qubit QCBMs trained with the Bell and Sparse targets exhibit overparameterization when $p>2$, 3- and 4-qubit QCBMs trained with the GHZ target exhibit overparameterization when $p<20$, yet in Fig. \ref{fig:variance_gradient} there is no indication that these models will train efficiently, based on the gradient at initialization.  Likewise the variance of 6- and 8-qubit QCBMs saturates (as predicted in \cite{McClean2018}).  This can indicate that the circuit forms a 2-design, yet in the main text we show that with the GHZ target these models can find low loss solutions.

\subsection*{Hessian Spectrum}
Our final use characterize the landscape of the QCBM where gradient-based training converges.  In the previous section the variance of the gradient is used to diagnose barren plateaus, and uses a first order (gradient) approximation that most directions explored by gradient descent are locally flat.  The Hessian matrix is a second order method that characterizes the local curvature around critical points (where $\nabla \mathcal{L}\to 0$). The Hessian matrix $\mathcal{H}_{ij} = \nabla_{\theta_i}\nabla_{\theta_j}(\mathcal{L})$ is a square matrix that provides important information about the local curvature near a critical point and has been used as a metric to understand the classical neural network landscape \cite{sagun2016eigenvalues} as also the quantum neural network landscape \cite{huembeli2021characterizing}.
\begin{figure}[htbp]
    \centering
    \includegraphics[width=.98\linewidth]{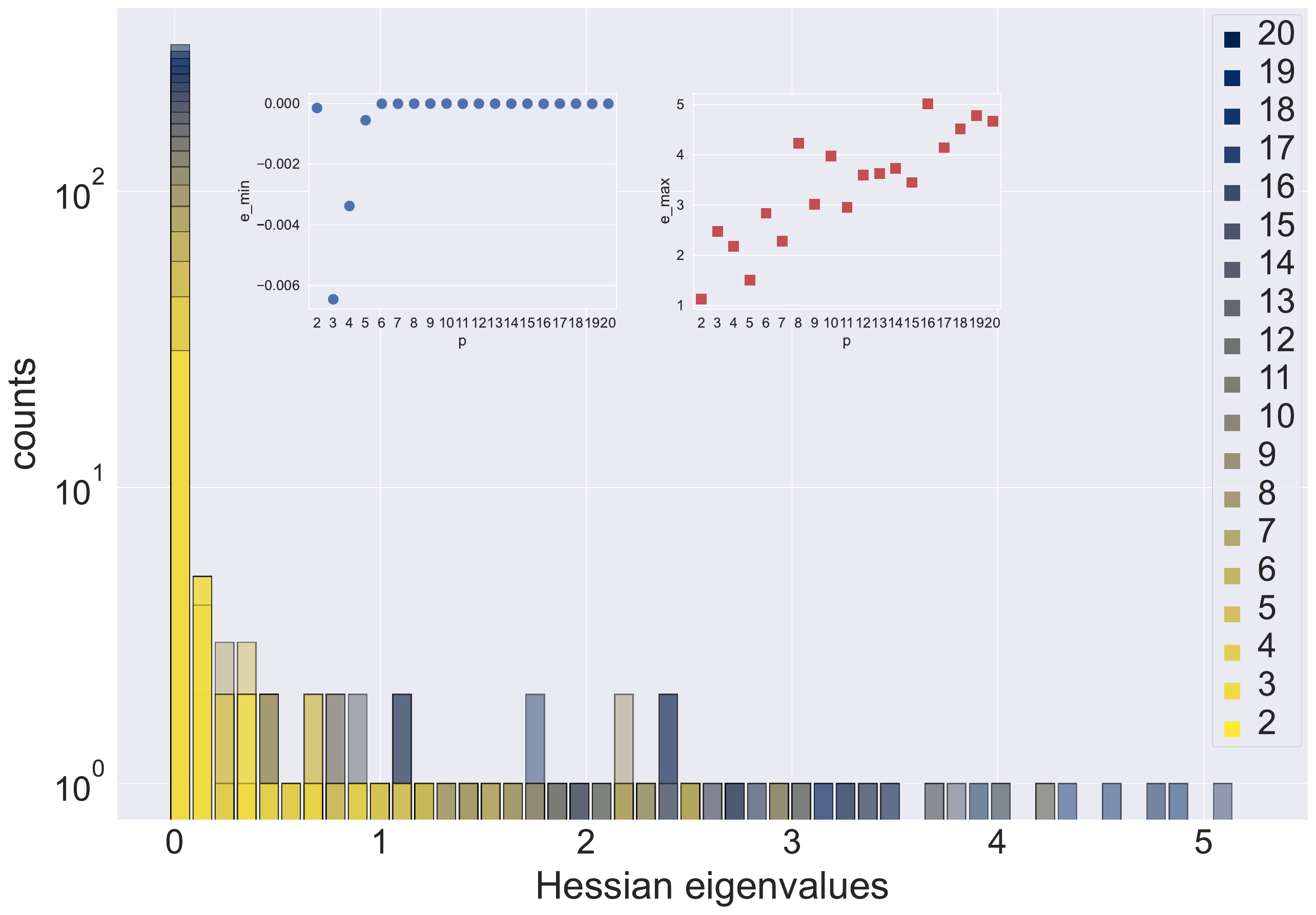}
    \caption{Eigenvalue spectra of Hessian matrix of 4-qubit QCBMs constructed with the HEA. Inset plots show the minimum (e\_min)and maximum eigenvalue (e\_max) at each depth.}
    \label{fig:4qhessian}
\end{figure}

We use the final trained parameters (where the training has converged after 200 optimization steps) and compute the Hessian to characterize the local curvature. By computing the eigenvalues and eigenvectors of the Hessian matrix, we can determine the nature of the critical point. If all eigenvalues are positive (negative), the critical point is a local minimum (maximum) of the function.  If the eigenvalues have mixed signs, the critical point is a saddle point.
\begin{figure}[htbp]
    \centering
    \includegraphics[width=.98\linewidth]{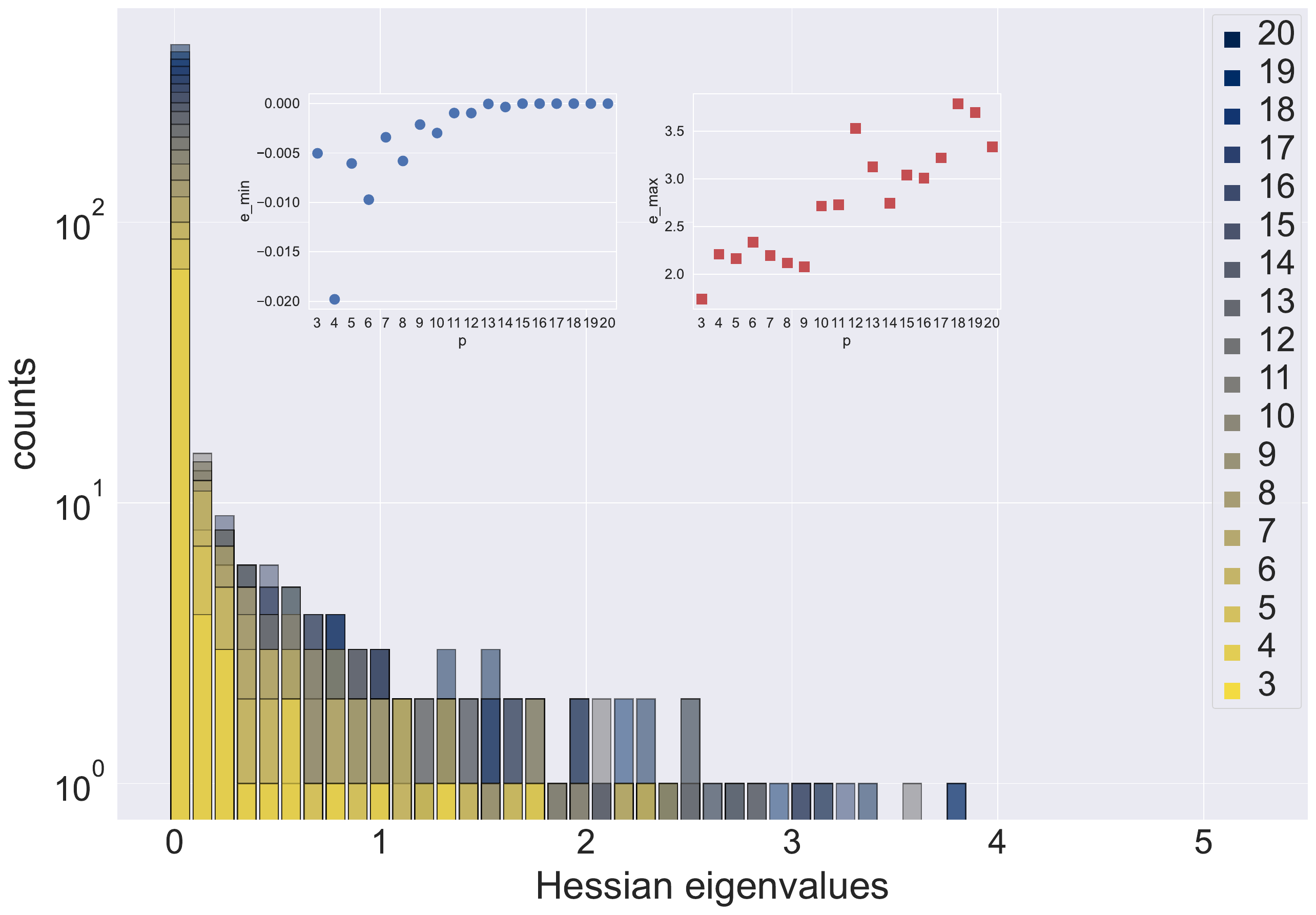}
    \caption{Eigenvalue spectra of Hessian matrix of QCBMs constructed with the HEA using 6 qubits.Inset plots show the minimum (e\_min)and maximum eigenvalue (e\_max) at each depth.}
    \label{fig:6qhessian}
\end{figure}

As seen in Figs. \ref{fig:4qhessian},\ref{fig:6qhessian} the QCBM spectra is dominated by a large number of zero eigenvalues--degenerate or flat curvature.  Similar results have been observed in classical neural networks, where the Hessian eigenvalues contain a bulk centered at zero, and a discrete subset of non-zero values \cite{sagun2016eigenvalues}.  For supervised models \cite{sagun2016eigenvalues} the number of non-zero eigenvalues is proportional to the number of classes.  For the QCBM Hessian spectra we do not count the number of non-zero eigenvalues, but in Figs. \ref{fig:4qhessian}, \ref{fig:6qhessian} we plot the maximum and minimum eigenvalues.  These values show that at small to moderate depths, the Hessian spectra show small negative curvature, indicating that the training converges to a saddle point. As the depth of QCBM increase, the critical points become minima with large degenerate spaces.
\end{document}